\documentclass[10pt,conference]{IEEEtran}
\IEEEoverridecommandlockouts
\usepackage{cite}
\usepackage{amsmath,amssymb,amsfonts}
\usepackage{algorithm}
\usepackage[noend]{algpseudocode}
\usepackage{graphicx}
\usepackage{textcomp}
\usepackage{xcolor}
\usepackage{listings}
\usepackage{alloy}
\usepackage{subfig}
\usepackage{booktabs}
\usepackage{soul}
\usepackage{tcolorbox}
\def\BibTeX{{\rm B\kern-.05em{\sc i\kern-.025em b}\kern-.08em
    T\kern-.1667em\lower.7ex\hbox{E}\kern-.125emX}}

\newcommand\alloymax{$\textnormal{Alloy}^{\textnormal{Max}}$}

\newtheorem{problem}{Problem}
\newtheorem{theorem}{Theorem}[section]
\newcommand{\transpose}{\sim\!\!}
\newcommand{\toolname}{\textsc{ATLAS}}

\soulregister\cite7
\soulregister\ref7
\newcommand\remove[1]{} 
\newcommand\edit[1]{#1} 
\newcommand\camera[1]{#1}

\begin{document}

\title{Constrained LTL Specification Learning from Examples}

\makeatletter
\newcommand{\linebreakand}{%
  \end{@IEEEauthorhalign}
  \hfill\mbox{}\par
  \mbox{}\hfill\begin{@IEEEauthorhalign}
}
\makeatother

\author{
\IEEEauthorblockN{Changjian Zhang}
\IEEEauthorblockA{\textit{Carnegie Mellon University} \\
Pittsburgh, PA USA \\
changjiz@andrew.cmu.edu}
\and
\IEEEauthorblockN{Parv Kapoor}
\IEEEauthorblockA{\textit{Carnegie Mellon University} \\
Pittsburgh, PA USA\\
parvk@andrew.cmu.edu}
\and
\IEEEauthorblockN{Ian Dardik}
\IEEEauthorblockA{\textit{Carnegie Mellon University} \\
Pittsburgh, PA USA\\
idardik@andrew.cmu.edu}
\and
\IEEEauthorblockN{Leyi Cui}
\IEEEauthorblockA{\textit{Columbia University} \\
New York, NY USA\\
lc3542@columbia.edu}
\linebreakand
\IEEEauthorblockN{R\^omulo Meira-G\'oes}
\IEEEauthorblockA{\textit{The Pennsylvania State University} \\
State College, PA USA \\
romulo@psu.edu}
\and
\IEEEauthorblockN{David Garlan}
\IEEEauthorblockA{\textit{Carnegie Mellon University} \\
Pittsburgh, PA USA \\
dg4d@andrew.cmu.edu}
\and
\IEEEauthorblockN{Eunsuk Kang}
\IEEEauthorblockA{\textit{Carnegie Mellon University} \\
Pittsburgh, PA USA \\
eunsukk@andrew.cmu.edu}
}

\maketitle

\thispagestyle{plain}
\pagestyle{plain}

\begin{abstract}
Temporal logic specifications play an important role in a wide range of software analysis tasks, such as model checking, automated synthesis, program comprehension, and runtime monitoring. Given a set of positive and negative examples, specified as traces, \emph{LTL learning} is the problem of synthesizing a specification, in \emph{linear temporal logic (LTL)}, that evaluates to true over the positive traces and false over the negative ones. In this paper, we propose a new type of LTL learning problem called \emph{constrained LTL learning}, where the user, in addition to positive and negative examples, is given an option to specify one or more \emph{constraints} over the properties of the LTL formula to be learned. We demonstrate that the ability to specify these additional constraints significantly increases the range of applications for LTL learning, and also allows efficient generation of LTL formulas that satisfy certain desirable properties (such as minimality). We propose an approach for solving the constrained LTL learning problem through an encoding in first-order relational logic and reduction to an instance of the \emph{maximal satisfiability (MaxSAT)} problem. An experimental evaluation demonstrates that \toolname{}, an implementation of our proposed approach, is able to solve new types of learning problems while performing better than or competitively with the state-of-the-art tools in LTL learning. 
\end{abstract}


\section{Introduction}
\emph{Temporal logic (TL)} specifications are a class of specification notations that are used to specify how a system behaves over time. TL specifications, such as linear temporal logic (LTL)~\cite{ltl}, have been used in a wide range of software analysis tasks, such as model checking~\cite{clarke2018handbook}, reactive synthesis~\cite{PnueliR89}, program understanding~\cite{Lemieux2015-ltl-mining}, and runtime monitoring~\cite{BauerLS11}. Despite its utility, formalizing and specifying desired system properties in temporal logic is a notoriously challenging and  error-prone process~\cite{Zeller2011-spec-mining,Rozier2016-spec-bottleneck,Lutz2023-LTL-Sketch}.


One active area of research that has the potential to overcome this challenge is \emph{specification learning}, where the goal is to automatically infer formal specifications from example traces \cite{Lemieux2015-ltl-mining,Neider2018-Learning,Lutz2023-LTL-Sketch,Zeller2011-spec-mining,Camacho2019LearningIM}. These traces may be derived from system executions or manually created by developers as examples.
The extracted specification could then be used by developers to comprehend and debug the behavior of a system, document its expected properties, or perform verification through methods such as model checking and theorem proving~\cite{Zeller2011-spec-mining,Ammons2002-mining,Ernst07-mining}.

In this work, we specifically focus on the problem of \emph{LTL learning from examples}; that is, given a set of \emph{positive} and \emph{negative} traces,
generate an LTL specification that evaluates to \emph{true} over the positive traces and \emph{false} over the negative ones. Several prior works have investigated this problem \cite{Neider2018-Learning,Lutz2023-LTL-Sketch,Li2011-ltl-mining,Lemieux2015-ltl-mining}. One of the challenges in these existing approaches is providing \emph{control} over the generated LTL \remove{expression}\edit{formula}. Typically, a given set of (positive and negative) traces are only a \emph{partial} specification of the underlying system behavior to be captured by the resulting LTL specification. \edit{A formula that is learned solely from such traces may fail to precisely capture the system behavior and need to be further refined.}

\remove{For example, consider a robotic system that is programmed to move around and visit two regions, \textsf{A} and \textsf{B}, while avoiding a dangerous area \textsf{R}. To infer an LTL specification for the robot, the engineer may provide the following two example traces: $\langle \textsf{A}, \textsf{B}, \textsf{B}, \ldots \rangle$ (positive, representing the robot visiting region \textsf{A} followed by \textsf{B}) and $\langle \textsf{R}, \textsf{R}, \ldots \rangle$ (negative, representing the undesirable behavior where the robot enters \textsf{R}). The following is an LTL specification that may be learned from these two traces:
$\textsf{A} \land \mathbf{X}(\mathbf{G}\textsf{B})$}
\remove{That is, the robot must start in region \textsf{A} and then immediately move to and stay in \textsf{B}. Technically, this expression is a solution to the LTL learning problem, but it does not precisely capture the behavior of the system. An ideal LTL specification would constrain the robot from ever moving into region \textsf{R}, while also not requiring the robot to start in region \textsf{A}.}

\edit{For example, consider a mobile navigation robot that moves around three regions, \textsf{A}, \textsf{B}, and \textsf{R} (the last one being a dangerous area that the robot should avoid). Suppose that the robot developer wishes to use formal specifications to precisely document the robot behavior and to leverage them for verification. An LTL learning tool can be used to extract such specifications from traces representing sample behaviors of the robot. Specifically, consider the following two sample traces collected from an operation log: $\langle \textsf{A}, \textsf{B}, \textsf{B}, \ldots \rangle$ (positive; the robot visiting \textsf{A} and then \textsf{B}) and $\langle \textsf{B}, \textsf{R}, \textsf{R}, \ldots \rangle$ (negative; the robot enters \textsf{R} and never leaves, indicating a potentially unsafe behavior). A specification that may be learned from these traces is: $\textsf{A} \land \mathbf{X}(\mathbf{G}\textsf{B})$, i.e., the robot should visit \textsf{A} initially and then stay in \textsf{B}. Technically, this formula is a valid solution to the learning problem. However, it also fails to capture the developer's intent, i.e., to prohibit the unsafe scenario. Instead, a more ideal specification would constrain the robot from being stuck at \textsf{R}, while not requiring it to start in region \textsf{A}.}

\edit{Devising an additional set of traces to cover a more diverse set of system behaviors may be one way to overcome this issue. However, in general, determining a set of traces that precisely capture the desired characteristics of a specification is a challenging task. For example, one may wish to extract a specification that fits into a certain pattern or a class (e.g., a safety or liveness property), minimize the number of times that a certain proposition appears, or learn a formula that is close to another specification (e.g., in the context of specification repair). LTL learning approaches that rely solely on examples lack the expressive power to allow fine-grained control over the characteristics of the formula to be inferred.}

\remove{To provide greater control over the properties of LTL specifications to be synthesized from examples,}\edit{To this end}, we propose the \emph{constrained LTL learning} problem as a generalization over the original LTL learning problem. The key idea is to allow the user to specify, besides the set of positive and negative traces, additional \emph{constraints} that must hold over the \emph{syntactic structure} of the generated LTL formula. These constraints, based on first-order logic (FOL), can be used to express a wide range of properties about LTL expressions, thus giving the user more control over the solution space being explored. \remove{For example, the engineer may specify constraints stating that ``the resulting LTL expression must mention proposition \textsf{R}'' and ``$\textsf{A} \lor \textsf{B}$ must hold in the initial state''.}\edit{For example, one may specify a constraint stating that ``the resulting LTL formula must mention region \textsf{R}'' or ``the formula should follow the pattern of $\mathbf{G}\phi$''.}
In addition, the user can specify \emph{optimization objectives} to maximize or minimize the syntactic structure in certain ways, e.g., minimize the formula size or maximize the similarity against another LTL formula. Together, these constraints and objectives can help bias generated solutions towards more desirable ones.

We present a constrained LTL learning tool, named \toolname{}, that solves this problem by encoding it into \alloymax{} \cite{alloy,Alloy-Max}, an extension of Alloy, a specification language based on first-order relational logic. A problem in \alloymax{} is eventually reduced to an instance of the \emph{maximal satisfiability (MaxSAT)} problem \cite{handbook-maxsat}. We show that  constrained LTL learning is a simple but powerful generalization of the original LTL learning problem, enabling a new set of use cases beyond those that were possible before in the existing state-of-the-art tools.
In addition to its applicability, our experimental evaluation shows that our approach to solving  constrained learning problems performs significantly better than the existing methods over a set of benchmark problems, and that it performs competitively over unconstrained problems.  

\edit{The primary target users of \toolname{} are \textbf{expert specifiers} and \textbf{tool builders}. For experts in formal logic who currently use LTL learning tools, \toolname{} provides them with a more powerful way to control the learning process to extract specifications that precisely capture their intent. Moreover, the generality of FOL means that \toolname{} has the potential to serve as an underlying, general-purpose engine for other specification-based tools that rely on LTL learning (e.g., specification weakening or repair, as described in Section \ref{sec:case-studies}). In addition, a tool built upon \toolname{} could automatically convert domain-specific constraints or requirements in natural language to FOL, enabling \textbf{non-expert users} to also benefit from LTL learning.}

The contributions of this paper are as follows:
\begin{itemize}
    \item The formulation of the \emph{constrained LTL learning} problem, a generalized learning problem that allows users to define FOL constraints and optimization objectives over the  structure of the learned formula (Section \ref{sec:problem}).
    \item An encoding of the learning problem in \alloymax{}, which is then solved as a MaxSAT problem (Section \ref{sec:encoding} and \ref{sec:learning-approach}), and a tool implementing the proposed technique, named \toolname{}.
    \item A set of case studies from different application domains (Section~\ref{sec:case-studies}), and a set of benchmark results demonstrating the performance of our approach on both constrained and unconstrained learning problems (Section \ref{sec:eval}).
\end{itemize}
\begin{figure}[!t]
    \centering
    \includegraphics[width=0.7\linewidth]{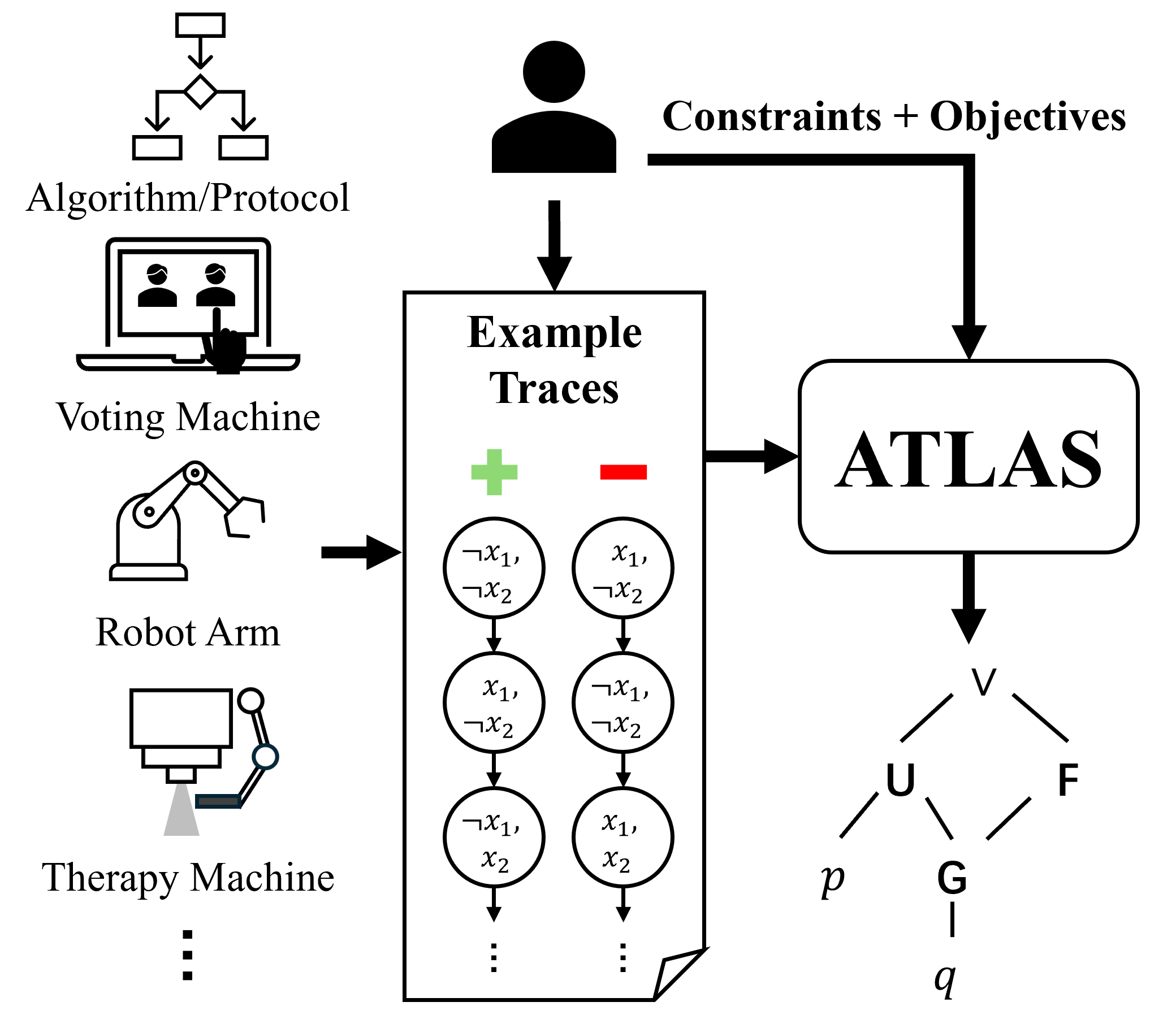}    
    \caption{\small{Overview of a constrained LTL learning problem in \toolname{}.}
    }
    \label{fig:overview}
    \vspace{-10pt}
\end{figure}

\section{Motivating Example}\label{sec:motivation}

In this section, we demonstrate the \remove{necessity}\edit{need} for more general and customizable constraints and objectives for LTL learning, \remove{using a case study from the domain of specification mining \cite{Ammons2002-mining,Dallmeier2010-test-mining,Peng23-mining,Ernst07-mining,Zeller2011-spec-mining}.}\edit{using a case study from specification mining \cite{Ammons2002-mining,Dallmeier2010-test-mining,Peng23-mining,Ernst07-mining,Zeller2011-spec-mining}, a major application of LTL learning.}

\edit{Consider a system analyst working on analyzing a multi-processing algorithm, developed by other engineers. 
The analyst plans to employ formal verification but the algorithm is designed without a formal specification. This lack of specification is common in practice; even in safety-critical domains, many systems do not have formally defined requirements until a later testing phase, when the need to support rigorous verification arises~\cite{Rozier2016-spec-bottleneck}. Although the analyst has a background in formal logic, without an in-depth understanding of the algorithm, it would be challenging to extract a specification that is not only consistent with the algorithm but also representative of its behavior. Specification mining is one approach to overcome this bottleneck---by generating executions (e.g., from an operation log), one can use LTL learning to infer specifications that capture the system behavior. These specifications can then be used for documentation, regression testing, or formal verification \cite{Rozier2016-spec-bottleneck,Zeller2011-spec-mining}. }


\begin{figure}[!ht]
    \centering
    
    \begin{algorithmic}[1]
        \Procedure{MultiProcessingAlgo}{$i$}
        \While{true}
        \State $a1_i$: $skip$ \normalfont{// Non-critical section}
        \State $a2_i$: $flag_i \leftarrow true$
        \State $a3_i$: $\textbf{await} \ (flag_{1-i} = false)$
        \State $cs_i$: $skip$ \normalfont{// Critical section}
        \State $a4_i$: $flag_i \leftarrow false$
        \EndWhile
        \EndProcedure
    \end{algorithmic}
    
    
    \caption{A multi-processing algorithm under analysis.}
    \label{fig:peterson}
    \vspace{-5pt}
\end{figure}

\remove{We illustrate this workflow using our given example.}\edit{We illustrate this workflow using a sample algorithm shown in Figure \ref{fig:peterson}; here, $i$ is the identifier  (ID) for a process, and only two concurrent processes are considered (ID = 0 or 1).}
To generate  execution traces, \edit{the implementation is instrumented with labels on instructions.}
\remove{let}\edit{Let} each label in Figure \ref{fig:peterson} (e.g., $a2_i$ on line 4) be a proposition that is true right before process $i$ executes that line of code, and becomes false after the line is executed. For instance, $a2_i$ becomes true right before the instruction $flag_i \leftarrow true$ is executed and becomes false after this assignment. \edit{Then, each trace is a sequence of \emph{pairs} of propositions, where each pair describes the current locations of the concurrent processes}. Figure \ref{fig:peterson-pos} shows an example of \remove{a system trace}\edit{such traces}.


\begin{figure}[!t]
    \centering
    \includegraphics[width=0.75\linewidth]{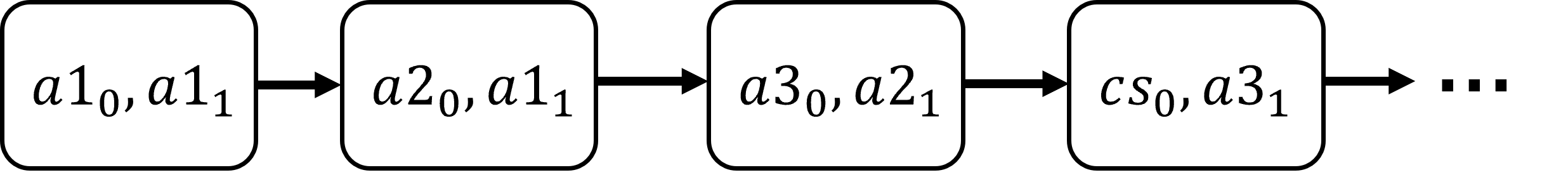}
    \caption{\remove{A positive example trace generated from the algorithm in Figure \ref{fig:peterson}. A box indicates a state. The variables in a box indicate the true propositions in that state.}\edit{A positive example trace of the algorithm in Figure \ref{fig:peterson}.}}
    \label{fig:peterson-pos}
    \vspace{-10pt}
\end{figure}

\paragraph{\edit{Limitations of existing methods}} \edit{The analyst may use an LTL learning tool to infer a specification from  traces. When an existing learner such as Flie \cite{Neider2018-Learning} is applied to the trace in Figure \ref{fig:peterson-pos},
one possible output formula is $a1_0$. This formula is technically valid, as it is consistent with this particular execution, but is also arguably not descriptive of the overall algorithm behavior. Providing more traces may not yield a different solution, as every possible trace starts with a state where $a1_0$ is true.}

\remove{In this case, specifying constraints over the LTL formula is required.
One way is to provide a template for safety invariants $\mathbf{G}\phi$ (i.e., $\phi$ always holds) in an LTL learner with template support such as \cite{Lutz2023-LTL-Sketch}.
However, without further constraining $\phi$ (e.g., $\phi$ must include $cs_i$), it can possibly return $\mathbf{G}\mathbf{F}(a1_0)$.
It is a valid liveness property but not a safety invariant and again does not reflect mutual exclusion. Additional constraints and objectives are needed to mine the accurate safety invariant. A high-level description of these can be of the form:}

\edit{One approach for learning more meaningful formulas is to allow the user to constrain the space of candidate formulas to be explored. For instance, suppose that the analyst wants to learn a safety invariant of form $\mathbf{G}\phi$ (i.e., $\phi$ always holds). One way to achieve this goal is to use a learner that takes a specification \emph{template} as an additional input (such as LTLSketcher \cite{Lutz2023-LTL-Sketch}). Although templates are a significant improvement, they also fall short of providing fine-grained control that may be desired. For example, without further constraining $\phi$ (e.g., $\phi$ should not contain temporal operators), a template-based learner may return $\mathbf{G}\mathbf{F}(a1_0)$, which is a valid liveness property but not a safety invariant as intended by the analyst.}

\paragraph{\edit{Proposed approach}}\edit{We argue that additional constraints and objectives are needed to express the analyst's preference on what a meaningful safety invariant looks like. For example, consider the following requirement:}
\begin{quote}
    ``Learn an LTL formula \remove{in \emph{the form of $\mathbf{G}\phi$} and \emph{maximize} the use of $cs_0$ and $cs_1$ in $\phi$}\edit{that  is in form $\mathbf{G}\phi$ where $\phi$ does not contain temporal operators, and  \emph{maximizes} the use of $cs_0$ and $cs_1$ in $\phi$}''.
\end{quote}
\remove{By specifying this requirement with our approach, the engineer can find a formula $\mathbf{G}(\neg (cs_0 \land cs_1))$ indicating the mutual exclusion property. However, such objectives cannot be expressed using existing tools.}

\edit{It first defines a constraint that the formula needs to be a safety invariant. Then, it defines an objective to maximize the use of certain propositions that the analyst is  interested in (i.e., the critical sections in the code). As explained in Section \ref{sec:problem}, it can be expressed as constraints in FOL, and our approach, \toolname{}, can find formula $\mathbf{G}(\neg (cs_0 \land cs_1))$, which captures a \emph{mutual exclusion} property. Such constraints and objectives cannot be expressed using the existing learning tools.}

\begin{figure}[!ht]
    \centering
    \includegraphics[width=0.75\linewidth]{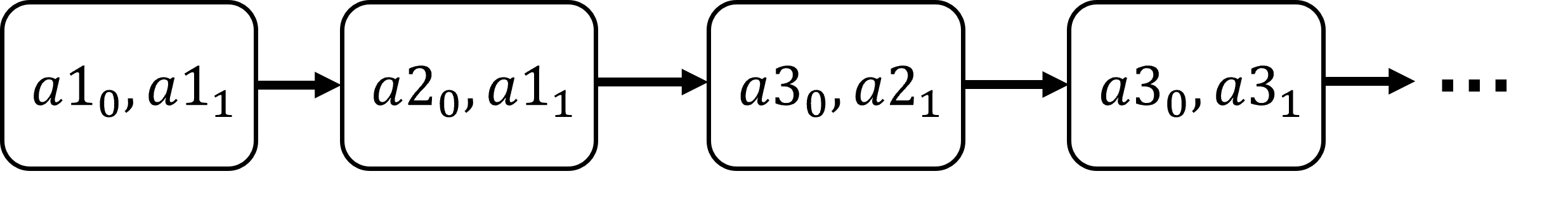}
    \caption{A negative example trace generated from the algorithm.}
    \label{fig:peterson-neg}
\end{figure}

\remove{To further demonstrate why such constraints and objectives are useful, consider the case where the algorithm generates another execution trace as shown in Figure \ref{fig:peterson-neg}. Since the execution is stuck at state $(a3_0, a3_1)$ where both processes are waiting for the other to falsify the $flag_i$ variables. It indicates a deadlock which is not desirable. In this case, the engineer decides to learn a property that can rule out this behavior. The engineer might specify this through a common liveness pattern $\mathbf{G}(\phi \rightarrow \mathbf{F}\psi)$, which can be expressed by a tool like \cite{Lutz2023-LTL-Sketch}. However, without additional constraints, it may return $\mathbf{G}(a3_0 \rightarrow \mathbf{F}a1_0)$. This is a valid property that ensures no deadlock; however, it is unintuitive because it requires a process to enter $a3_0$ before $a1_0$, even though $a1_0$ occurs before $a3_0$ in the algorithm's procedure. The following is a high-level requirement for improving the quality of the formula:}

\edit{As another example, consider the trace shown in Figure \ref{fig:peterson-neg}. This trace is considered negative, as it depicts an undesirable  behavior where the execution is stuck at state $(a3_0, a3_1)$ (i.e., both processes are waiting for the other to reset the flag). In this case, the analyst wants to learn a liveness property that would rule out this behavior, in  form of $\mathbf{G}(\phi \rightarrow \mathbf{F}\psi)$. Again, given this pattern as a template, a tool like LTLSketcher \cite{Lutz2023-LTL-Sketch} may return $\mathbf{G}(a3_0 \rightarrow \mathbf{F}a1_0)$. This is a valid   liveness property; however, it is also rather unintuitive as it requires a process to enter $a3_0$ before $a1_0$, even though $a1_0$ occurs before $a3_0$ in the algorithm's procedure.} 

\edit{To further improve this formula, the analyst using \toolname{} can specify a constraint capturing the following requirement:}
\begin{quote}
    ``$\psi$ should be a label after $\phi$ as they appear in the algorithm procedure\remove{, e.g., if $\phi = a3_0$, then $\psi$ should only be either $cs_0$ or $a4_0$}''.
\end{quote}
\remove{Finally, the engineer learns a more comprehensible liveness property $\mathbf{G}(a3_0 \rightarrow \mathbf{F} cs_0)$. It manifests as an implicit program requirement (deadlock-freedom) that the engineer failed to consider initially.}\edit{This constraint requires that, for example, if $\phi = a3_0$, then $\psi$ should  be either $cs_0$ or $a4_0$. Given this constraint as an input, \toolname{} generates formula $\mathbf{G}(a3_0 \rightarrow \mathbf{F} cs_0)$, which is arguably more intuitive than the above one. This formula corresponds to a \emph{deadlock-free} property.}

\paragraph{\edit{Summary}} \edit{The algorithm described in Figure \ref{fig:peterson} is derived from the well-known Peterson's Mutual Exclusion Algorithm \cite{Peterson:1981}, with a deadlock defect introduced for illustration. The learned formulas in the example might look relatively simple to specify. However, even for a simple algorithm as this one, existing learning techniques often fail to return solutions that precisely capture the system behavior.
This limitation cannot be addressed only by adding more traces, and there is a need for more expressive constraints and objectives.}

To our knowledge, \toolname{} is the first LTL learning tool that is expressive enough to encode each of the constraints in the example above. It offers users the capability to interactively and gradually add custom constraints and objectives  to extract valuable specifications from examples.  \remove{Other than mining from existing code, we illustrate our tool's extended expressiveness and flexibility for learning using three case studies from different domains (Section \ref{sec:case-studies}).}\edit{Later in (Section \ref{sec:case-studies}), we demonstrate the expressive power and generality of \toolname{} through additional 
use cases beside specification mining, such as specification repair and invariant weakening.}

\section{Preliminaries}

\subsection{Linear Temporal Logic}
Linear Temporal Logic (LTL) \cite{ltl} is an extension of propositional logic with temporal operators. Its syntax is as follows:
\begin{align*}
    \phi ~:=~ & p ~|~ \neg \phi ~|~ \phi \land \psi ~|~ \phi \lor \psi ~|~ \phi \rightarrow \psi ~|~ \\
            & \mathbf{G} \phi ~|~ \mathbf{F} \phi ~|~ \mathbf{X} \phi ~|~ \phi \mathbf{U} \psi
\end{align*}
where $p \in AP$ is an atomic proposition of a finite set of propositions $AP$. An LTL formula is interpreted over an infinite trace $\sigma \in (2^{AP})^\omega$. Specifically, the temporal operators are interpreted as:
\begin{itemize}
    \item $\mathbf{G} \phi$ (Globally): $\phi$ holds in all future states.
    \item $\mathbf{F} \phi$ (Finally): $\phi$ holds eventually in some future state.
    \item $\mathbf{X} \phi$ (Next): $\phi$ holds in the next state.
    \item $\phi \mathbf{U} \psi$ (Until): $\phi$ always holds until $\psi$ becomes true in some future state.
\end{itemize}

\subsection{LTL Learning from Examples}
In a typical setting, \emph{LTL learning} defines the problem of inferring an LTL formula from positive and negative example traces \cite{Neider2018-Learning}: Given a set of atomic propositions $AP$, let $P, N \subset (2^{AP})^\omega$ be two \edit{(potentially empty)} disjoint sets of infinite traces, where $P$ are \emph{positive examples} and $N$ are \emph{negative examples}. We call $\mathcal{S} = (P, N)$ a \emph{sample}. Then, the task of LTL learning is to find a formula $\phi$ such that $\forall \sigma \in P:$ the trace $\sigma$ \emph{satisfies} formula $\phi$, $\sigma \models \phi$, and $\forall \bar\sigma \in N:$ the trace $\sigma$ \emph{does not satisfy} formula $\phi$, $\bar\sigma \not\models \phi$. There exists a trivial solution to separate $P$ and $N$ in the form $\bigvee_{a \in P} \bigwedge_{b \in N} \varphi_{a,b}$, where $\varphi_{a,b}$ separates each example pair $(a,b)$. However, this solution is obviously over-fitting and less helpful in practice. Thus, we are often interested in finding an LTL formula of \emph{minimal} size.


\subsection{Alloy and Alloy$^{Max}$}
Alloy \cite{alloy} is a modeling language based on first-order relational logic with transitive closure. With its SAT-based engine, the Alloy Analyzer has been applied to a wide range of problems, including protocol verification~\cite{zave17, Trippel2018}, test generation~\cite{khurshid-04}, and bug finding~\cite{dennis-issta06, jackson-issta00}. 

\alloymax{} \cite{Alloy-Max} is an extension of Alloy with a capability to express and analyze problems with optimal solutions. It introduces a small addition of language constructs to specify problems with optimality as an objective and translation from an \alloymax{} problem to a \emph{maximum satisfiability (MaxSAT)} problem \cite{handbook-maxsat}, which can be solved by a MaxSAT solver. Specifically, \alloymax{} can be used to: (1) \emph{maximize or minimize relations}, e.g., maximizing allowed packets while adhering to network security policies \cite{Narain05}; (2) \emph{define soft constraints}, e.g., adding ``participants' time preferences'' in meeting scheduling; and (3) \emph{define priorities for various objectives}, e.g., prioritizing morning meeting times over afternoon ones in meeting scheduling.

Interested readers are encouraged to refer to Alloy \cite{alloy} and \alloymax{} \cite{Alloy-Max} for more details about their capabilities.

\section{Problem Formulation}\label{sec:problem}
In this work, we propose a new type of LTL learning problem called the \textit{constrained LTL learning problem}.

\subsection{Constrained LTL Learning} 
\begin{problem}\label{prob:ltl-learning}
    A constrained LTL learning problem is defined as a tuple $\langle AP, \mathcal{S}, \Phi, \Psi \rangle$ where
    \begin{itemize}
        \item $AP$ is a \emph{finite} set of atomic propositions,
        \item $\mathcal{S} = (P, N)$ is a sample,
        \item $\Phi$ is a first-order predicate that constraints the \emph{syntactic structure} of the learned formula, and
        \item $\Psi$ is an optimization objective over the syntactic structure of the formula.
    \end{itemize}
    The goal of the problem is to find an LTL formula $\phi$ such that $\forall \sigma \in P: \sigma \models \phi$, $\forall \bar\sigma \in N: \bar\sigma \not\models \phi$, $\Phi\bigl(\textsf{syntax}(\phi)\bigr)$ holds, and $\textsf{syntax}(\phi)$ optimizes $\Psi$, where $\textsf{syntax}(\phi)$ represents the syntactic structure of $\phi$.
\end{problem}

For the syntactic constraint $\Phi$, a user could specify it to learn an invariant $\mathbf{G} \phi$ where $\phi$ is a propositional formula,
a liveness property $\mathbf{G} (\phi \rightarrow \mathbf{F} \psi)$, or a GR(1) formula, which is widely used in reactive synthesis \cite{GR(1)-synthesis}. Many of these cannot be expressed by existing tools.
Moreover, for the objective $\Psi$, a user could specify it to optimize the syntactic structure in certain way, e.g., minimizing the size of the formula.


\subsection{Syntactic Constraint $\Phi$}\label{sec:problem-constraint} 
We formally define the constructs for the syntactic constraint $\Phi$.
For an LTL formula $\phi$, $\textsf{syntax}(\phi) = \langle \mathcal{L}, \mathcal{R}, root \rangle$ represents its syntax tree where $\mathcal{L}, \mathcal{R} \subseteq N \times N$
are the left child and right child relations, respectively; $N$ is the set of nodes in the syntax tree, and $root$ is the root node. In particular, $N = \bigcup_{op \in \mathcal{N}} N_{op}$ where $\mathcal{N} = \{\mathbf{G,F,U,X},\land,\lor,\rightarrow,\neg,AP\}$ and $N_{op}$ is the set of nodes for a particular operator type or atomic propositions.

\begin{figure}[!t]
\begin{align*}
& \Phi ::=~ \varphi ~|~ \text{funcDef} \\
& \text{funcDef} ::=~ \textbf{func}~\text{identifier} (\text{var}) = \text{expr} \\
& \varphi ::=~ \text{elementary} ~|~ \neg \varphi ~|~ \varphi \land \varphi ~|~ \varphi \lor \varphi ~|~ \varphi \rightarrow \varphi ~|~ \varphi \leftrightarrow \varphi ~|~ \\
& \quad\quad\quad \forall~\text{varDecl}: \varphi ~|~ \exists~\text{varDecl}: \varphi \\
& \text{elementary} ::=~ \text{expr} \in \text{expr} ~|~ \text{expr} \subseteq \text{expr} ~|~ \text{expr} = \text{expr} ~|~ \\
& \qquad\qquad\qquad\;\; |\text{expr}|~\text{compOp}~\text{number} \\
& \text{expr} ::= ~ \text{const} ~|~ \text{var} ~|~ \text{comprehension} ~|~ \text{expr} \cup \text{expr} ~|~ \\
& \qquad\quad\quad \text{expr} \cap \text{expr} ~|~ \text{expr} \setminus \text{expr} ~|~ \text{expr} \times \text{expr} ~|~ \\
& \qquad\quad\quad \text{expr}~\textbf{.}~\text{expr} ~|~  ^\wedge\text{expr} ~|~ *\text{expr} ~|~ \transpose~\text{expr} ~|~ \text{func} \\
& \text{comprehension} ::=~ \{\text{var}~|~\varphi\} \\
& \text{func} ::=~ \text{identifier}(\text{expr}) \\
& \text{varDecl} ::=~ \text{var} \in \text{expr} \\
& \text{compOp} ::=~ = ~|~ < ~|~ > ~|~ \leq ~|~ \geq \\
& \text{const} ::=~ \text{identifier} ~|~ \emptyset \\
& \text{var} ::=~ \text{identifier} ~|~ (\text{identifier}[, \text{identifier}]*)
\end{align*}
\caption{Abstract syntax of syntactic constraint $\Phi$.}
\label{fig:constraint-syntax}
\vspace{-10pt}
\end{figure}

Then, $\Phi$ is a FOL constraint over $\textsf{syntax}(\phi)$. Figure \ref{fig:constraint-syntax} shows its abstract syntax, which is an extension to FOL and set theory with the following operations for improved expressiveness:
\begin{itemize}
    \item $s.r = \{b~|~a \in s \land (a, b) \in r\}$, join of set $s \subseteq N$ and relation $r \subseteq N \times N$. Also, for simplicity, let $a.r$ = $\{a\}.r$ where $a \in N$.
    \item $r_1.r_2 = \{(a, c)|(a, b) \in r_1 \land (b, c) \in r_2\}$, join of relations. 
    \item $^\wedge(r) = r \cup r.r \cup r.r.r \cup \ldots$, transitive closure of a relation $r \subseteq N \times N$.
    \item $*(r) = {^\wedge}(r) \cup \{(a, a)~|~a \in N\}$, reflexive transitive closure of a relation $r \subseteq N \times N$.
    \item $\transpose(r) = \{(b, a)~|~(a, b) \in r\}$, inverse of a relation. 
\end{itemize}
Moreover, a user can define functions to reuse common expressions. By default, we introduce the following functions:
\begin{itemize}
    \item $l(n) = n.\mathcal{L}$, left child of a node $n$.
    \item $r(n) = n.\mathcal{R}$, right child of a node $n$.
    \item $\textsf{desc}(n) = n.{^\wedge}(\mathcal{L} \cup \mathcal{R})$, all descendent nodes of a node $n$ by using transitive closure.
    \item $\textsf{subNodes}(n) = n.{*}(\mathcal{L} \cup \mathcal{R})$, all descendent nodes of a node $n$ including itself.
\end{itemize}

\noindent\textbf{Example 1}: To learn a liveness property $\mathbf{G} (\phi \rightarrow \mathbf{F} \psi)$ where $\phi$ and $\psi$ are propositional formulas, the user can define the following constraints:
\begin{align}
    & n_{\mathbf{G}} \in N_{\mathbf{G}} ~\land~ n_{\rightarrow} \in N_{\rightarrow} ~\land~ n_{\mathbf{F}} \in N_{\mathbf{F}} \\
    & root = n_{\mathbf{G}} ~\land~ l(root) = n_{\rightarrow} ~\land~ r(n_{\rightarrow}) = n_{\mathbf{F}} \\
    & \textsf{subNodes}\bigl(l(n_{\rightarrow})\bigr) \cap N_{\{\mathbf{G,F,U,X}\}} = \emptyset \\
    & \textsf{desc}(n_{\mathbf{F}}) \cap N_{\{\mathbf{G,F,U,X}\}} = \emptyset
\end{align}
where $N_{\{\mathbf{G,F,U,X}\}} = \bigcup_{op \in \{\mathbf{G,F,U,X}\}} N_{op}$ and lines are connected by $\land$. Specifically, line (1) declares instances of nodes, line (2) defines the structure $\mathbf{G(\phi \rightarrow \mathbf{F} \psi)}$, and lines (3) and (4) define $\phi$ and $\psi$ should not contain temporal operators.

\subsection{Optimization Objective $\Psi$}\label{sec:problem-objective}
The syntax for optimization objective $\Psi$ is an extension to the syntax defined in Figure \ref{fig:constraint-syntax}:
$$\Psi :=~ min^{[k]}(\text{expr}) ~|~ max^{[k]}(\text{expr}) ~|~ \text{expr} \approx^{[k]} \emptyset ~|~ \textbf{[} \varphi \textbf{]}^{[k]}$$
Specifically, we introduce the following operators:
\begin{itemize}
    \item $min(s)$, where $s$ is a non-empty set and the number of elements in $s$ should be minimized.
    \item $max(s)$, where $s$ is a non-empty set and the number of elements in $s$ should be maximized.
    \item $s \approx \emptyset$, minimizes set $s$ and ideally makes it empty.
    \item $[\varphi]$, where $\varphi$ is a constraint and is optional to be true.
\end{itemize}
An optimization operator can be assigned a superscript $k \in \mathbb{N}^+$ (e.g., $min^k(s)$) indicating its priority. An objective with a higher priority should be optimized before all the other objectives with lower priorities. When the superscript $k$ is omitted, the constraint has the lowest priority $k = 1$.

\noindent\textbf{Example 2:} Since there exists an over-fitting solution for any learning problem, we often prefer to learn a formula that is minimal in its size. This can be expressed as:
\begin{equation}
    \mathcal{L} \cup \mathcal{R} \approx \emptyset
\end{equation}
It minimizes the size of the child relations (edges in a syntax tree), which, in turn, minimizes the size of the formula.

\noindent\textbf{Example 3:} In addition, when a user wants to learn a formula that maximizes the use of a set of critical propositions $critical \subseteq AP$, they can define:
\begin{equation}
    max^2\bigl(\textsf{subNodes}(root) \cap N_{critical}\bigr)
\end{equation}
The use of $max^2$ specifies that this goal should be optimized with priority 2, which happens before minimizing the total size of the solution as shown in the previous example.

\section{Technical Approach}\label{sec:encoding}

We reduce a constrained LTL learning problem to an instance of relational model finding. In particular, we use \alloymax{}, which allows us to define and solve problems in first-order relational logic with the support for all the operators for syntactic constraint $\Phi$
and optimization objective $\Psi$. 

The idea of our encoding is inspired by the SAT encoding proposed by Neider and Gavran \cite{Neider2018-Learning}, which is inspired by bounded model checking \cite{Handbook-model-checking-SAT}.
The encoding assumes that all traces in the sample set are ultimately periodic, known as \emph{lasso traces}. Such a trace can be represented as $uv^\omega$ where $u \in (2^{AP})^*$ and $v \in (2^{AP})^+$. The observation is that, given a finite set of $AP$, the set of states visited by $uv^\omega$ (starting from any time point $t$) is finite and can be determined based only on the finite prefix $uv$. Therefore, this enables us to check LTL temporal operators w.r.t. infinite traces \cite{Neider2018-Learning}.
In addition, the user provides the maximum number of sub-formulas to bound the search space.
Furthermore, we employ several heuristics, described in later sections, to make our \alloymax{} encoding more succinct and efficient.

The encoding can be divided into six parts: LTL syntax encoding, LTL semantics encoding, problem-specific encoding, learning objective encoding, custom syntactic constraints, and optimization objectives.

\subsection{Syntax Encoding}
\label{sec:syntax-encoding}
We model the syntax of an LTL formula as a directed acyclic graph (DAG). Essentially, a syntax DAG is a syntax tree with shared common sub-formulas. This helps reduce the search space.
The following code snippet shows how this syntax graph can be modeled in \alloymax{}:
\begin{alloy}
// A "sig" (signature) defines a type of atoms.
abstract sig DAGNode {
 // fields of a signature become relations
 l: set DAGNode, r: set DAGNode
}
// "fact" defines a block of constraints.
fact { all n: DAGNode | n not in n.^(l + r) }
// "extends" defines sub-types of a signature.
sig And, Or, Imply, Until extends DAGNode {} {
 // "one" means exactly one element.
 one l and one r
}
sig Neg, F, G, X extends DAGNode {} {
 // "no" means empty set.
 one l and no r
}
abstract sig Literal extends DAGNode {} {
  no l and no r
}
one sig LearnedLTL { Root: DAGNode }
fun root: one DAGNode { LearnedLTL.Root }
\end{alloy}
Line 2 defines the parent signature for all nodes in a DAG, and line 4 defines the left/right child relations, $l$ and $r$. Line 7 constrains each node to have no paths to itself,
i.e., the graph should be acyclic. Specifically, this is achieved by computing all the descendent nodes of $n$ through transitive closure $^\wedge(l + r)$ as defined in Section \ref{sec:problem-constraint}, where $+$ is the union operator.

Lines 9-19 define the sub-types of nodes including operators and atomic propositions. Specifically, we add constraints for: binary operators to have exactly one left and one right child (lines 9-12), unary operators to have only left child (line 13-16), and atomic propositions to have no children (line 17-19). Finally, lines 20-21 define a helper function for accessing the root node. Therefore, relation $l$ and $r$, and function $root$ construct the syntax of the learned formula, $\textsf{syntax}(\phi) = \langle \mathcal{L, R}, root \rangle$.

\subsection{Semantics Encoding}\label{sec:encoding-operator}
This section describes the encoding for the semantics of LTL. We show only the definitions for $\lor, \mathbf{X}, \mathbf{F},~\text{and}~\mathbf{U}$. Encodings for the other operators can be defined in a similar manner.
\begin{alloy}
abstract sig SeqIdx {}
abstract sig Trace {
 // relation, Trace x SeqIdx x SeqIdx
 lasso: SeqIdx -> SeqIdx,
 // relation, Trace x DAGNode x SeqIdx
 val: DAGNode -> SeqIdx
}
fun seqIndices[t: Trace]: set SeqIdx { .. }
fun futureIdx[t:Trace,i:SeqIdx]: set SeqIdx {..}
\end{alloy}
Line 1 defines signature \textsf{SeqIdx} for time points. Lines 2-7 define the signature for lasso traces. Specifically, for any trace $t=uv^\omega$, it has a $lasso$ relation $(t, i, i')$ where $i = |uv| - 1$ is the ending index of its prefix $uv$ and $i' = |u|$ is the starting index of loop $v$. It also has a $val$ relation, where any $(t, n, i) \in val$ stands for: the trace $t$ satisfies the sub-formula, represented by the sub-DAG from node $n$, starting from time $i$.

Lines 8-9 define two helper functions \textsf{seqIndices} and \textsf{futureIdx}. 
Their functionalities are:
\begin{itemize}
    \item $\textsf{seqIndices}(t)$ returns all the time points of a trace $t$, as the example traces can be in different lengths;
    \item $\textsf{futureIdx}(t, i)$ returns all the future time points of a trace $t$ starting from time $i$ (including $i$).
\end{itemize}
They are computed using the $lasso$ relation and a relation $next \subseteq \textsf{SeqIdx} \times \textsf{SeqIdx}$ that defines a total order over the time points, connecting a time $i$ with its next time point $i+1$.

For instance, Figure \ref{fig:lasso-trace} shows an example lasso trace $uv^\omega$. The solid black arrows represent relation $next = \{(0, 1), (1, 2)\}$, and the dashed arrow represents relation $lasso = \{(2, 1)\}$ of this trace. Based on the $next$ relation, for time $i = 0$ and $i = 1$, the next time points are $i' = 1$ and $i' = 2$, respectively. For time $i = 2$, the state value at $i'= 3$ equals to the value at $i' = 1$ because of the repeating $v$'s, and $i' = 1$ can be retrieved by the $lasso$ relation. Therefore, $\{(2, 3), (3, 4), \ldots\}$ are unnecessary tuples (shown in Figure \ref{fig:lasso-trace}), and we only need to model the finite prefix $uv$ of a lasso trace.


\begin{figure}[!ht]
    \centering
    \includegraphics[width=0.85\linewidth]{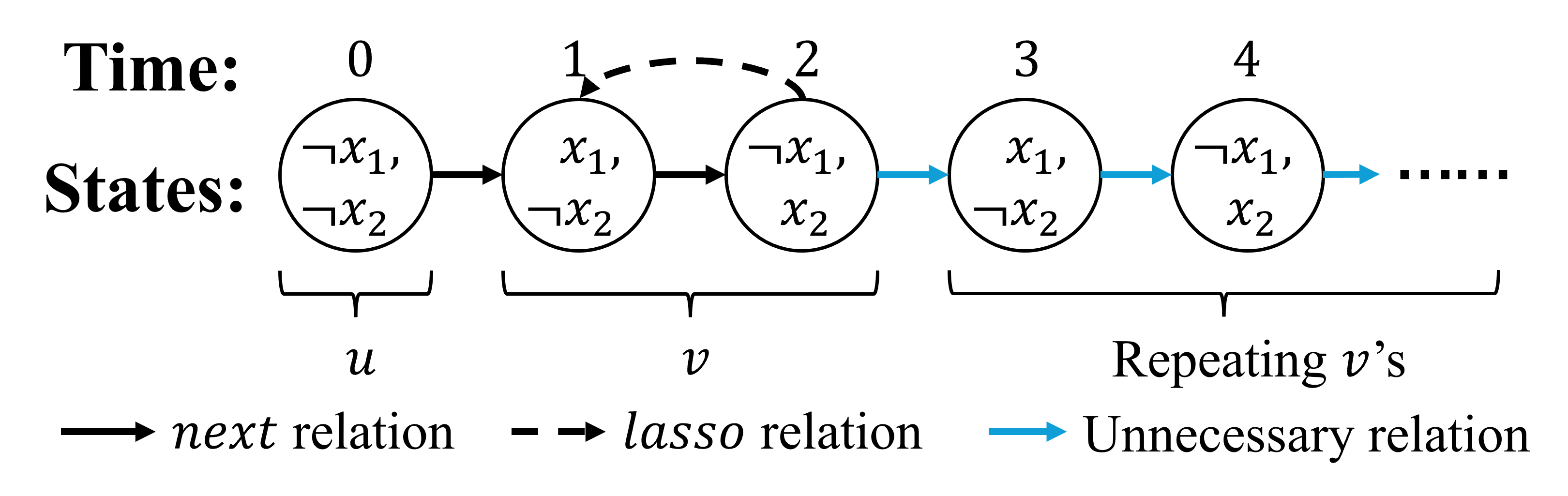}
    \caption{An example lasso trace with two atomic propositions.}
    \label{fig:lasso-trace}
\end{figure}

For this trace, function $\textsf{seqIndices}(t) = \{0, 1, 2\}$, function $\textsf{futureIdx}(t, 0) = \{0, 1, 2\}$, and function $\textsf{futureIdx}(t, 2) = \{2, 1\}$. A heuristic is applied that uses transitive closures to compute them, e.g., $i.*(next+t.lasso)$ returns all future time points of $i$ without comparing the order of time points.

\subsubsection{$\lor$-operator}
For any node $n \in N_{\lor}$ and time $i$ being in the time range of $uv$ (computed by \textsf{seqIndices}), the tuple $(t, n, i)$ is in $val$ if and only if, at time $i$, its left sub-formula $n.l$ or right sub-formula $n.r$ is in $val$, i.e., one of the sub-formulas holds.
\begin{alloy}
all t: Trace, n: Or, i: seqIndices[t] |
 n->i in t.val iff
  (n.l->i in t.val or n.r->i in t.val)
\end{alloy}

\subsubsection{$\mathbf{X}$-operator}
For any node $n \in N_{\mathbf{X}}$ and time $i$, $(t, n, i) \in val~\text{if and only if}~(t, n.l, i') \in val$, where $i'$ is the next time point of $i$ computed by expression $i.(next + t.lasso)$. For example in Figure \ref{fig:lasso-trace}, $(next + t.lasso) = \{(0, 1), (1, 2), (2, 1)\}$. Thus, $i.(next + t.lasso)$ returns the next time point of time $i$, considering the loop $v$ in $t = uv^\omega$.

\begin{alloy}
all t: Trace, n: X, i: seqIndices[t] |
n->i in t.val iff n.l->i.(next+t.lasso) in t.val
\end{alloy}

\subsubsection{$\mathbf{F}$-operator}
For any node $n \in N_{\mathbf{F}}$ and time $i$, $(t, n, i) \in val~\text{if and only if}~\exists i' \in \textsf{futureIdx}(t, i) : (t, n.l, i') \in val$, where operator \A{some} stands for $\exists$ and \textsf{futureIdx} computes all the future time points.
For example, for the trace in Figure \ref{fig:lasso-trace}, when $i = 0$, $\textsf{futureIdx}(t, i) = \{0, 1, 2\}$. It means that from $i = 0$, all the possible state values can be represented by states at $i' = 0, 1, 2$. Thus, we assert that there exists a time point in these future indices where $p$ holds.
\begin{alloy}
all t: Trace, n: F, i: seqIndices[t] |
 n->i in t.val iff
  some i': futureIdx[t, i] | n.l->i' in t.val
\end{alloy}

\subsubsection{$\mathbf{U}$-operator}
Finally, the following code defines the $q \mathbf{U} p$ operator. Specifically, we leverage the heuristic from the SAT encoding of the $\mathbf{U}$-operator used in bounded model checking \cite{Handbook-model-checking-SAT}. It unfolds $\mathbf{U}$ by using the $\mathbf{X}$-operator in the sense that $q \mathbf{U} p = p \lor (q \land \mathbf{X}(q \mathbf{U} p))$, as defined in lines 4-5. In addition, on line 3, we constrain that sub-formula $p$ (i.e., $n.r$) will eventually be true from time $i$; otherwise, a trace where $p$ never becomes true would also satisfy this encoding.
\begin{alloy}
all t: Trace, n: Until, i: seqIndices[t] |
 n->i in t.val iff {
  some i': futureIdx[t, i] | n.r->i' in t.val
  n.r->i in t.val or (n.l->i in t.val and
   n->i.(next+t.lasso) in t.val)
 }
\end{alloy}

\subsection{Problem-Specific Encoding}
The following code snippet shows the \alloymax{} template for the problem-specific encoding, particularly the values of all the positive and negative traces.
\begin{alloy}[mathescape=true]
one sig /* $p \in AP$ */ extends Literal {}
one sig /* $i \in [0, max(|t|)), t\in \mathcal{S}$*/ extends SeqIdx {}
fact {
  first = /* $i_0$ */
  next = /* $\{(i_0, i_1), (i_1,i_2), \ldots\}$ */
}
abstract sig PositiveTrace extends Trace {}
abstract sig NegativeTrace extends Trace {}
one sig /*$t = uv^\omega \in P$*/ extends PositiveTrace {} {
  lasso = /* $(i_{|uv|-1}, i_{|u|})$ */
  /* $\{ (t, n, i)~|~n \in AP \land n \in uv(i) \}$ */ in val
  // & means set intersection.
  no /* $\{ (t, n, i)~|~n \in AP \land n \notin uv(i) \}$ */ & val
}
one sig /*$t = uv^\omega \in N$*/ extends NegativeTrace {} {
  // same above
}
\end{alloy}
In this template, the mathematical expressions in the comments \A{/*..*/} will be replaced by the actual parameters of a specific learning problem $\langle AP, \mathcal{S}, \Phi, \Psi \rangle$. Line 1 defines all the atomic propositions in $AP$ as literal DAG nodes. Lines 2-6 define the finite set of time points given the maximum length of traces in the sample $\mathcal{S}$ and explicitly specify the initial time $i_0$ and the $next$ relation. This leverages a heuristic in Alloy to improve performance through \emph{partial instances}~\cite{torlak2006kodkod}.

Lines 7-8 further divide the \textsf{Trace} signature into two sets, \textsf{PositiveTrace} and \textsf{NegativeTrace}. Then, lines 9-17 explicitly specify the value for a trace $t = uv^\omega$ in set $P$ and $N$.
Specifically, on line 10, its $lasso$ relation maps its last time index $i_{|uv|-1}$ to the start of the loop $i_{|u|}$. Lines 11-13 specify that for any atomic proposition node $n \in AP$: (1) if $n \in uv(i)$, where $uv(i)$ is the set of true propositions at time $i$ of prefix $uv$, then the tuple $(t, n, i)$ is in $val$ relation; (2) otherwise, it is not. E.g., for the trace $t$ in Figure \ref{fig:lasso-trace}, we have:
\begin{align*}
    \{ (t, x_1, i_1), & (t, x_2, i_2) \} \subseteq val~\text{and} \\
    \{ (t, x_1, i_0), (t, x_2, i_0), & (t, x_2, i_1), (t, x_1, i_2) \} \cap val = \emptyset
\end{align*}

\subsection{Learning Objective Encoding}
The following code allows \alloymax{} to generate an LTL formula that satisfies the sample.
\begin{alloy}
run {
 all t: PositiveTrace | root->T0 in t.val
 all t: NegativeTrace | root->T0 not in t.val
} for /* max number of DAG nodes */ DAGNode
\end{alloy}
Line 2 defines that for any positive trace $t \in P$, $(t, root, 0) \in val$, i.e., the learned formula is true on trace $t$ from time 0. In contrast, for any negative trace $t \in N$, $(t, root, 0)$ is not in $val$ (line 3). Finally, on line 4, the user provides the maximum number of DAG nodes allowed for a particular problem.

\subsection{Custom Structural Constraints}
\alloymax{} supports all the constructs for defining the structural constraint $\Phi$, including FOL, set operations, and the additional join, transitive closure, and inverse operations\footnote{Due to limited space, the  translation from \toolname{} constraints and objectives into \alloymax{} is omitted. The translation, however, is straightforward as Alloy itself is based on FOL and well-suited for encoding the constraints.}.
For example, to encode the constraints for $\mathbf{G}(\phi \rightarrow \mathbf{F} \psi)$ (Example 1 of Section \ref{sec:problem-constraint}), we have:
\begin{alloy}
fun desc[n: DAGNode] { n.^(l+r) }
fun subNodes[n: DAGNode] { n.*(l+r) }
one sig G0 extends G {}
one sig Imply0 extends Imply {}
one sig F0 extends F {}
fact {
 root = G0 and root.l = Imply0 and Imply0.r = F0
 no (G+F+Until+X) & subNodes[Imply0.l]
 no (G+F+Until+X) & desc[F0]
}
\end{alloy}
Lines 1-2 correspond to the \textsf{desc} and \textsf{subNodes} functions. Lines 3-5 declare instances for $N_{\mathbf{G}}, N_{\rightarrow}, N_{\mathbf{F}}$, respectively. Finally, lines 7-9 correspond to the constraints lines (2)-(4) in Example 1.

\subsection{Optimization Objective}
Similar to the syntactic constraints, \alloymax{} supports all the optimization operators for defining  objective $\Psi$. Specifically, it has the following mappings:
\begin{itemize}
    \item $min^k(s)$: \A{minsome[k] s}
    \item $max^k(s)$: \A{maxsome[k] s}
    \item $s \approx^k \emptyset$: \A{softno[k] s}
    \item $[f]^k$: \A{soft[k] fact}
\end{itemize}
where $k$ is an optional priority. For example, by default, we include the following optimization goal to minimize the size of the learned formula (Example~2 in Section~\ref{sec:problem-objective}): 
\begin{alloy}
softno l + r
\end{alloy}
Moreover, to maximize the use of critical propositions (Example~3), we have:
\begin{alloy}
maxsome[2] subNodes[root] & (p + q + r)
\end{alloy}
where let $critical = \{p, q, r\} \in AP$.

\section{Learning by \alloymax{}}\label{sec:learning-approach}
\subsection{Solving \alloymax{} with MaxSAT}
We briefly explain how Alloy and \alloymax{} solve a problem using a SAT and a MaxSAT solver, respectively; interested readers should refer to the original papers for more details \cite{Alloy-Max,daniel2000alloy,emina2007kodkod}. A relation in Alloy is translated into a matrix of Boolean variables -- each of which is true if and only if the tuple represented by this particular variable is in the relation; and a relational expression is represented by operations over one or more Boolean matrices.

For example, consider a relation $r: A \times B$ where $A = \{A1, A2\}$ and $B = \{B1, B2\}$. Then, this relation is represented by a set of Boolean variables $\{r_{11}, r_{12}, r_{21}, r_{22}\}$ where, for example, $r_{11}$ is true if and only if tuple $(A1, B1)$ is in $r$.
Then, in \alloymax{}, it encodes optimization goals as weighted soft clauses such that a MaxSAT solver finds optimal solutions by maximizing the total sum of weights. For example, the \alloymax{} operator \A{softno} minimizes the number of tuples in a relation, the expression \A{softno r} will be converted to:
$$(\neg r_{11})^{k} \land (\neg r_{12})^{k} \land (\neg r_{21})^{k} \land (\neg r_{22})^{k}$$
where, e.g., $(\neg r_{11})^{k}$ is a soft clause with weight $k$ that may or may not be satisfied. Thus, this formula finds a relation $r$ with a minimized number of tuples in it, and ideally, $r = \emptyset$.

In addition, Alloy supports blocking a solution and using an incremental solver to find a new solution. Thus, we leverage this feature to enumerate solutions of a learning problem.

\subsection{\edit{Correctness}}

\camera{Our approach is sound but complete only up to the user-provided bound on the maximum number of DAG nodes. A correctness proof can be found in the Appendix.}

\subsection{Quality of Solutions}\label{sec:solution-quality}
Since there can be multiple LTL formulas that satisfy an LTL learning problem, additional constraints and objectives are often necessary for finding ``useful'' solutions w.r.t. a problem domain. Moreover, enumerating solutions satisfying the constraints is also a critical functionality. Our approach supports all of them. Furthermore, other than problem-dependent constraints, additional sets of constraints are also useful for avoiding trivial or less-satisfactory solutions.
We list a few constraints that we find useful from our experience.

\emph{Disable reusing in DAG:} The DAG encoding was designed to improve performance. When minimizing against a DAG, it guarantees a minimal number of \emph{distinct} sub-formulas. However, this does not always produce a minimal LTL formula in its \emph{syntax tree size} where repeating sub-formulas are also counted. For example, consider $\mathbf{F} p \lor \mathbf{F} \mathbf{G} p$ and $\mathbf{F} \mathbf{G} p \lor \mathbf{F} \mathbf{G} p$. The latter one has a bigger syntax tree size but a smaller DAG size as the sub-formula $\mathbf{F} \mathbf{G} p$ is reused \cite{Lutz2023-LTL-Sketch}. Such behavior might reduce the usefulness of the tool for certain problems and can be disabled with the following constraint:
$$\forall n \in N \setminus N_{AP}: |n.{\sim}(\mathcal{L} \cup \mathcal{R})| \leq 1 \land \mathcal{L} \cap \mathcal{R} = \emptyset$$
where $n.{\sim}(\mathcal{L} \cup \mathcal{R})$ returns the parent nodes of $n$ (which could be more than one in a DAG).

\emph{Avoid tautology:} Another example is avoiding tautologies in solutions. This is often necessary when only positive traces are provided, where the learner may return a tautology ($true$) as a valid solution.
$$\forall n \in N_{\rightarrow}: l(n) \neq r(n)$$

\emph{Negation normal form (NNF):} The use of negation may result in some equivalent but less readable formulas, e.g., $\neg \mathbf{G}(x)$ versus $\mathbf{F}(\neg x)$ where the latter may be more readable and preferable. NNF is also helpful in automated theorem proving \cite{wos1992automated}.
Thus, we can enforce that negation should only be applied to atomic propositions.
$$\forall n \in N_{\neg}: l(n) \in N_{AP}$$

This is an inexhaustive list of problem-independent constraints that we find helpful.
Other examples include requiring $\phi$ to be in \emph{conjunction normal form} (CNF) or \emph{disjunction normal form} (DNF).

\section{Use Cases}
\label{sec:case-studies}
In this section, we present three use cases to demonstrate the need for syntactic constraints and/or optimization objectives. \edit{These use cases also show how tool builders can leverage \toolname{} as a back-end LTL learner to perform various specification-based tasks, such as specification repair and invariant weakening.}

\subsection{Specification Mining}

This use case demonstrates the usefulness of custom constraints and enumeration.
This case study is inspired by a voter fraud incident in Kentucky, USA, where corrupt officials manipulated a flaw in the system to ``flip'' votes \cite{fbi_2010}.
Figure \ref{fig:voting} shows an abstract model of the system as described in \cite{Romulo2023-safe-envelope}:
A voter enters the booth, enters a password, selects a candidate, votes, and finally confirms the vote before leaving the booth.


We generate example traces based on this model, manually check their correctness, and mark them as positive or negative. Specifically, a negative trace indicating a vulnerability of this system is: A voter leaves the booth before confirming their vote; then, a corrupt official enters the booth and uses the back button to select another candidate and confirms the choice.

\begin{figure}[!t]
    \centering
    \includegraphics[width=0.6\linewidth]{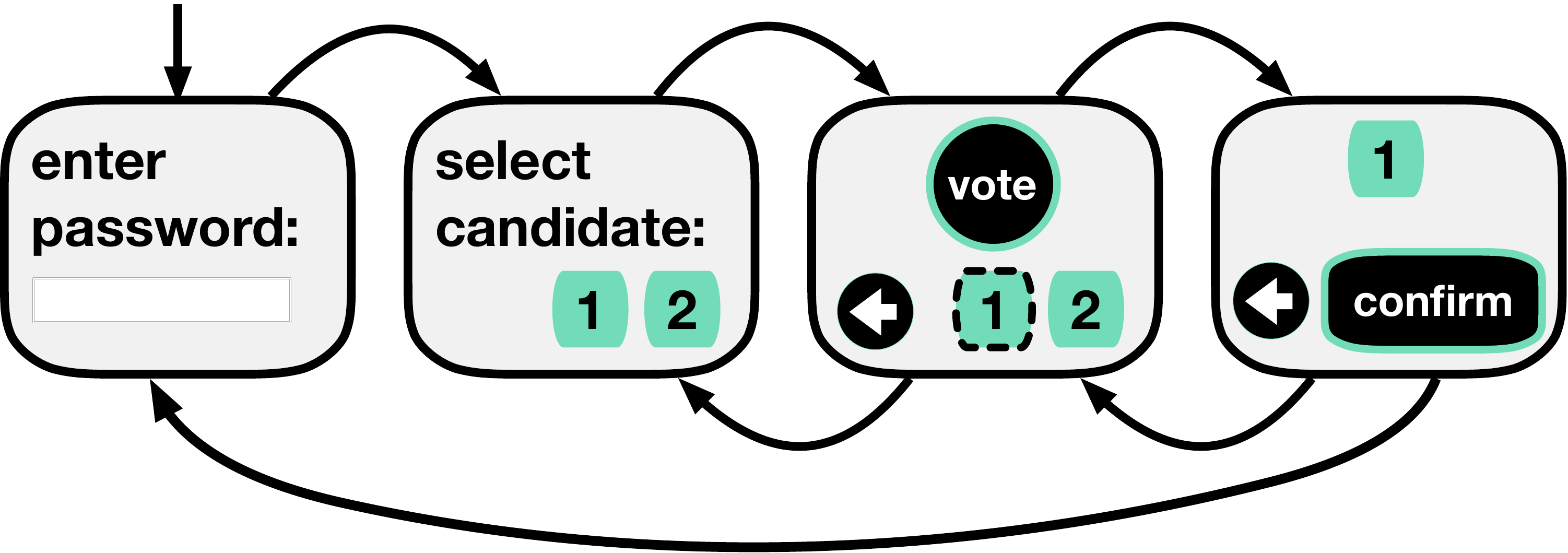}
    \caption{A state machine representing the voting machine \cite{Romulo2023-safe-envelope}.}
    \label{fig:voting}
    \vspace{-10pt}
\end{figure}

We want to learn a safety invariant to ensure election integrity. Particularly, the invariant should be of the form $\mathbf{G}\phi$, where $\phi$ should not contain any temporal operators. These constraints can be expressed as:
\begin{align}
    root \in N_{\mathbf{G}} \land l(root) \cap N_{\mathbf{G,F,U,X}} = \emptyset
\end{align}

In addition, our technique supports enumerating solutions that satisfy the given constraints, in the order of the formula size (or other custom optimization objectives). This is often helpful in finding variants of formulas. Specifically, in this case study, the enumeration helps us to find two safety invariants:
\begin{enumerate}
    \item $\mathbf{G}(\textsf{selectCandidate} \rightarrow \textsf{voterInBooth})$
    \item $\mathbf{G}(\textsf{officialInBooth} \rightarrow \textsf{enterPwd})$
\end{enumerate}
The first formula is in line with the key safety property defined by formal methods experts in \cite{Romulo2023-safe-envelope}. However, there is a
stronger, implicit requirement defined in \cite{Romulo2023-safe-envelope}: The election official is restricted from entering the booth after the voter enters her password. We observe that this requirement is captured precisely and explicitly by the second invariant.

\subsection{Specification Repair from Demonstrations}
In this use case, we demonstrate how custom optimization objectives can be helpful in learning. The use case is motivated by the field of autonomous agent interpretability using LTL specifications\cite{Shah2018BayesianIO}. 
Consider, for example, the setup in Figure \ref{fig:robarm} where the end effector of a robotic arm has to be actuated to perform some tasks. The designer provides an initial LTL specification: $\mathbf{F}(green) \land \mathbf{G}(\neg red)$;
i.e., reaching the green region and avoiding the red region.

However, after
a controller has been developed, the designer decides to refine the LTL specification to better reflect the system requirements. For example, consider the requirement that the robot should satisfy the above LTL formula but also visit the blue region in Figure~\ref{fig:robarm}.
The designer can generate a set of demonstrative traces and mark each one as positive or negative, depending on whether the robot successfully reaches the blue region in the trace. Moreover, the modification to the existing LTL should ideally be minimal while taking this additional requirement into account. We pose this as an LTL learning problem, where the associated objective is to minimize the \emph{edit distance}, i.e., minimizing the removal of existing sub-formulas as well as the addition of new sub-formulas.
The encoding of this minimal modification problem is defined as follows:
\begin{align}
& n_{\mathbf{G}} \in N_{\mathbf{G}} ~\land~ n_{\mathbf{F}} \in N_{\mathbf{F}} ~\land~ n_{\land} \in N_{\land} ~\land~ n_{\mathbf{\neg}} \in N_{\neg}  \\
& green, red \in  N_{AP} \\
& oldSpec =  \{(n_{\land}, n_{\mathbf{F}}), (n_{\land}, n_{\mathbf{G}}), (n_{\mathbf{F}}, green), \\
& \quad\quad\quad\quad\quad\nonumber (n_{\mathbf{G}}, n_{\neg}), (n_{\neg}, red)\} \\
& max^2\bigl((\mathcal{L} \cup \mathcal{R}) \cap oldSpec\bigr)
\end{align}
Lines (8)-(10) define the original formula. Line (11) minimizes the edit distance by using the $max$ operator (with priority 2). The learning process will first retain as many sub-formulas as possible from the old specification and then minimize the total size of the final formula.
With these constraints, our approach learns the ideal specification:
$\mathbf{F}(green) \land \mathbf{G}(\neg red) \land \mathbf{F}(blue)$,
where $blue$ stands for reaching the blue region.


\begin{figure}[!t]
    \centering
    \subfloat[A 7 DoF Panda Robotic Arm.]{
        \includegraphics[width=0.4\linewidth]{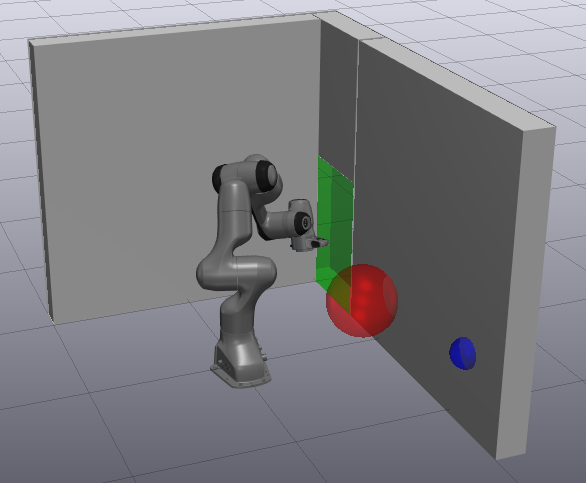}
        \label{fig:robarm}
    }
    \hfill
    \subfloat[A radiation therapy machine.]{
        \includegraphics[width=0.4\linewidth]{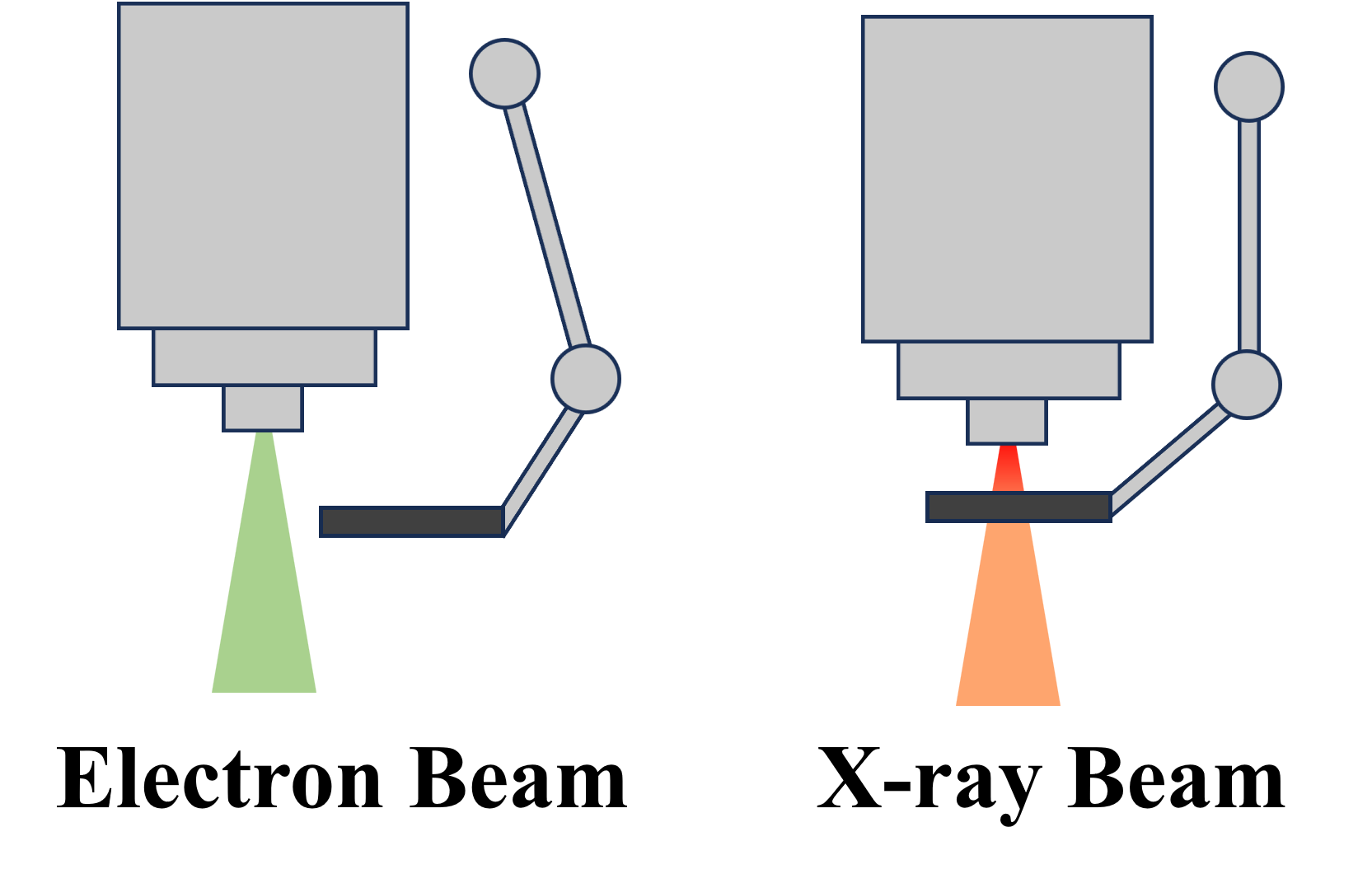}
        \label{fig:therac-25}
    }
    \caption{Examples of specification repair and weakening.}
    \label{fig:robarm-therac25}
    \vspace{-10pt}
\end{figure}

\subsection{Invariant Weakening}
In this case study, we show how complex LTL patterns (e.g., CNFs and DNFs) can be expressed using our technique. 
The idea of invariant weakening, or more generally, specification weakening, comes from requirements engineering \cite{Lamsweerde-conflict-goal,alrajeh-icse20}. As the environmental conditions for a software system may change over time and space, original requirements might become inadequate or inconsistent with the new environment, necessitating adaptation or weakening. This concept has been further explored in self-adaptive systems \cite{Whittle-RELAX-fuzzy,Sebastian2023-weakening,Chu-weakening}.


For instance, Figure \ref{fig:therac-25} is a radiation therapy machine similar to Therac-25 \cite{therac25} as described in \cite{CJ-Robustification}. The machine has two modes: Electron Beam mode and X-ray mode. A spreader must be inserted during the X-ray mode to attenuate the effect of the high-power X-ray beam and limit possible overdose. This safety requirement can be captured as the following invariant $P$:
$\mathbf{G}(\textsf{XrayMode} \rightarrow \textsf{SpreaderIn})$.

However, a system might become too restrictive to satisfy this safety property under certain environmental behavior. For example, when switching from X-ray to Electron beam, it might be acceptable that the spreader is out before the mode switching is completed, as long as the beam is not fired. Otherwise, we may have to disable all mode switching to ensure safety \cite{CJ-Robustification}. One way to mitigate this issue is to weaken the safety invariant (i.e., finding $P'$ s.t. $P \rightarrow P'$) from positive (acceptable) and negative (definitely unsafe) examples \cite{Sebastian2023-weakening}. An ideal weakened formula is: $\mathbf{G}(\textsf{XrayMode} \land \textsf{Fired} \rightarrow \textsf{SpreaderIn})$.


We consider a safety invariant of the form $\mathbf{G}(\phi \rightarrow \psi)$, where $\phi$ and $\psi$ are Boolean formulas. To achieve weakening, one can learn an invariant $\mathbf{G}(\phi \land \phi' \rightarrow \psi \lor \psi')$ where $\phi \land \phi'$ is a Boolean formula in CNF and $\psi \lor \psi'$ is a Boolean formula in DNF. In other words, an invariant can be weakened by (1) adding more conditions (as conjunctions) in the assumption, or (2) adding new acceptable conditions (as disjunctions) in the guarantee. This weakening objective can be expressed through the following constraints:
\begin{align}
& n_{\mathbf{G}} \in N_{\mathbf{G}} \land n_{\rightarrow} \in N_{\rightarrow} \land root = n_{\mathbf{G}} \land l(root) = n_{\rightarrow} \\
& \forall n \in \textsf{desc}(n_{\rightarrow}): n \in N_{\{\land,\lor,\neg,AP\}} \\
& \forall n \in \textsf{desc}(n_{\rightarrow}): n \in N_{\neg} \rightarrow l(n) \in N_{AP} \\
& \forall n \in \textsf{subNodes}\bigl(l(n_{\rightarrow})\bigr) \cap N_{\lor}: \textsf{desc}(n) \cap N_{\land} = \emptyset \\
& \forall n \in \textsf{subNodes}\bigl(r(n_{\rightarrow})\bigr) \cap N_{\land}: \textsf{desc}(n) \cap N_{\lor} = \emptyset \\
& l(n_{\rightarrow}) = \textsf{XrayMode} ~\lor~ \\
& \nonumber\qquad l(n_{\rightarrow}) \in N_{\land} \land l\bigl(l(n_{\rightarrow})\bigr) = \textsf{XrayMode} \\
& r(n_{\rightarrow}) = \textsf{SpreaderIn} ~\lor~ \\
& \nonumber\qquad r(n_{\rightarrow}) \in N_{\lor} \land l\bigl(r(n_{\rightarrow})\bigr) = \textsf{SpreaderIn}
\end{align}
Specifically, line (12) defines the structure $\mathbf{G}(\phi \rightarrow \psi)$. Line (13) stipulates that $\phi$ and $\psi$ contain only $\land,\lor,\neg$, and $AP$. Lines (14)-(16) state that $\phi$ is in CNF and $\psi$ is in DNF. Lines (17)-(18) state that $\phi$ may add conjuncts to \textsf{XrayMode}, and $\psi$ may add disjuncts to \textsf{SpreaderIn}. With these constraints, we can learn the ideal weakened formula as described above.

\section{Evaluation}\label{sec:eval}

We investigate the following research questions:
\begin{enumerate}
    \item \textbf{RQ1}: How does our approach perform compared to the state-of-the-art tool, Flie \cite{Neider2018-Learning}, over unconstrained LTL learning problems?  
    \item \textbf{RQ2}: For problems with structural constraints and/or optimization objectives as described in our case studies, how does our approach perform compared to the state-of-the-art tools?
\end{enumerate}
For RQ1, the metric we are concerned with is the total time taken to find the first satisfying formula. For RQ2, we evaluate performance through two key metrics: (1) the number of constrained problems solved and (2) the time taken to solve them. In particular, even though existing tools cannot encode constraints such as ours, they can be configured to enumerate all solutions until a solution satisfying the constraints is found; the goal here is to compare how long it takes for \toolname{} and the baselines to find an ideal solution.

We implemented \emph{\toolname} in Java and use OpenWBO \cite{open-wbo-sat14} as the MaxSAT solver for \alloymax{}. All experiments were run on a Linux machine with a 4-core 3.8GHz CPU and 8GB memory. For each problem, we impose a 180-second time out.

\subsection{Experimental Setup}
\edit{For fair comparison, we compare \toolname{} against the state-of-the-art tools satisfying the following two criteria: (1) no restrictions on the type of formula that can be learned and (2) the guarantee of finding minimal formulas.}

\subsubsection{RQ1}
\edit{For unconstrained problems, we compare our tool against the SAT-based algorithm of Flie \cite{Neider2018-Learning}. We did not use the decision-tree-based method of Flie as it does not provide the minimality guarantee.} \remove{To answer RQ1,}\edit{Then,} we leverage the same benchmark problems from Neider \cite{Neider2018-Learning} which are generated based on common LTL patterns \cite{Dwyer98-LTL-patterns}.
Specifically, this benchmark contains 485 problems with a number of examples ranging from 6 to 5000, a number of atomic propositions ranging from 2 to 9, and a length of traces ranging from 5 to 10.

\subsubsection{RQ2}
\edit{For constrained problems, we first compare our tool against Flie, which is configured to enumerate solutions until an expected formula is found. We also compare against LTLSketcher \cite{Lutz2023-LTL-Sketch} that can learn a formula given a user-defined template. It considers three types of placeholders in a template: (1) any LTL sub-formula, (2) any unary operator, and (3) any binary operator. However, it cannot express all our expected constraints and does not support enumeration either. Thus, we test whether it finds an expected formula in one run with the closest template regarding our required constraints. If it does not, we count it as a timeout. Also note that LTLSketcher is not used in RQ1 as it builds on Flie, and without user-provided templates, it reduces to Flie.}

\edit{Moreover, other tools are not considered because they do not meet our criteria. For example, SCARLET \cite{Raha_scalable_2022} is a performant learning tool but cannot handle Until and nested Eventually and Globally and also does not guarantee minimality.}

\remove{To answer RQ2,}\edit{Then,} we generate a set of constrained problems from our case studies with the expected solutions provided. \remove{We compare our tool against Flie \cite{Neider2018-Learning}, where Flie is configured to enumerate solutions until the expected formula is found. We also compare against LTLSketcher \cite{Lutz2023-LTL-Sketch} that can learn a formula given a user-defined template. It considers three types of placeholders in a template: (1) any LTL sub-formula, (2) any unary operator, and (3) any binary operator. However, it cannot express all our expected constraints and does not support enumeration either. Thus, we test whether it finds the expected formulas in one run with templates that closely capture our required constraints. If it does not find a solution satisfying our constraints, we count it as a timeout.}\edit{Specifically, we provide a set of solutions satisfying our constraints for each problem, and a problem is deemed solved when a tool can find any formula from the set.}

\noindent\textbf{Specification Mining.}
We generate problems from the two use cases described in Section \ref{sec:motivation} and \ref{sec:case-studies}. For Peterson's algorithm, we randomly generate 30 problems from a model specified in TLA+~\cite{lamport2002specifying}, with 12 to 16 example traces, 15 atomic propositions, and a trace length of 32. We add constraints for learning the mutual exclusion and the deadlock-free properties.
Then, when comparing against LTLSketcher, we use template $\mathbf{G}(?)$ for mutual exclusion and $\mathbf{G}(? \rightarrow \mathbf{F} ?)$ for deadlock-free, where $?$ stands for any sub-formula.

For the voting  example, we generate 10 problems with 8 to 19 example traces, 10 atomic propositions, and a trace length from 11 to 16. We add the constraint to learn an invariant in $\mathbf{G} \phi$ where $\phi$ is a propositional formula. However, the closest template in LTLSketcher is $\mathbf{G}(?)$ where $?$ can be any formula.

\noindent\textbf{Specification Repair.}
We generate 20 random problems from a robot arm simulator developed in \cite{kurtz2023temporal}. To simulate a specification repair process, we generate positive traces using the ideal specification and generate negative traces by negating a sub-formula in it. The problems have example traces ranging from 4 to 16, 3 atomic propositions, and a length of traces from 33 to 44. When comparing against LTLSketcher, since it has no way to express our optimization objective, we set an empty template to learn any formula.


\noindent\textbf{Invariant Weakening.}
We generate 78 random weakening problems for two types of weakening: let $AP = \{a, b, c\}$, (1) weaken $\mathbf{G}(a \rightarrow b)$ to $\mathbf{G}(a \land c \rightarrow b)$, and (2) weaken $\mathbf{G}(a \rightarrow b \land c)$ to $\mathbf{G}(a \rightarrow (b \land c) \lor b)$.
The number of examples of the problems ranges from 20 to 200, and the length of the examples ranges from 5 to 100. When comparing against LTLSketcher, the closest it can express is $\mathbf{G}(? \rightarrow ?)$.


\subsection{Results}

\begin{figure}[!t]
    \centering
    \includegraphics[width=\linewidth]{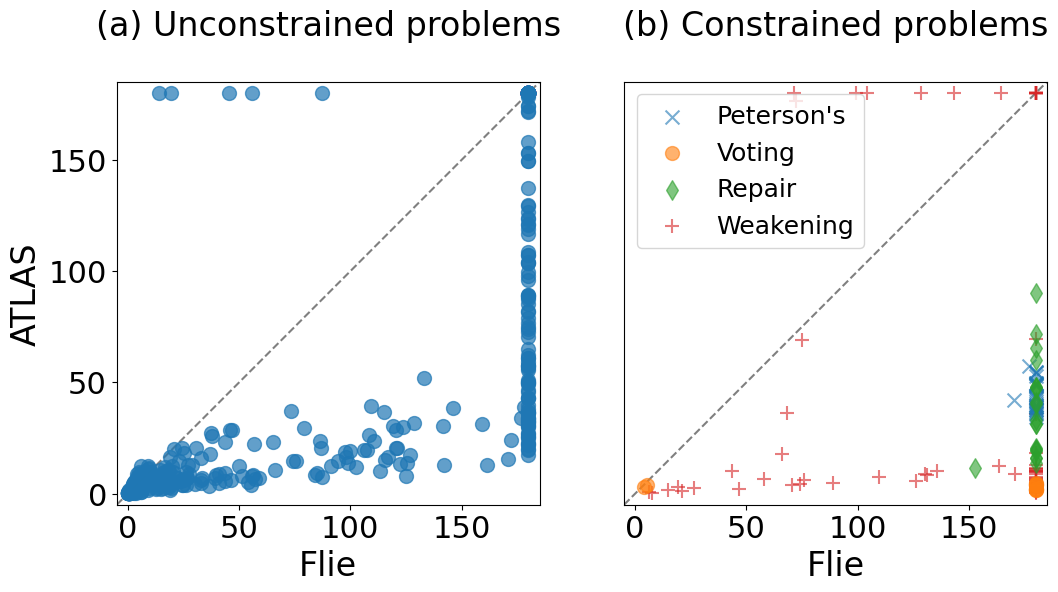}
    \caption{Comparisons of solving time in (a) unconstrained and (b) constrained problems. The axes indicate the time in seconds to solve a problem with a 180s timeout. A marker $(x, y)$ indicates a problem is solved by Flie in $x$ seconds and by \toolname{} in $y$ seconds. A dot below the diagonal line ($y=x$) indicates our tool is faster than Flie.}
    \label{fig:comparison}
    \vspace{-10pt}
\end{figure} 


\edit{All these tools take a file containing the example traces of a problem as input (with additional constraints or templates for RQ2). We checked the consistency of the learned formulas with our predefined solutions. Then, we report on the comparison results of \toolname{} against other tools.}

\noindent\textbf{RQ1.}
Figure \ref{fig:comparison}a shows the evaluation results for unconstrained problems. A blue marker below the diagonal line indicates that our approach is faster than Flie for a given problem. We find that our approach is faster than or equal to Flie in most of the problems  ($\mathbf{417/485 \approx 86\%}$), including where both timeout. The average solving time (including the 180s timeout) is: $\mathbf{96.83s}$ for Flie and $\mathbf{66.76s}$ for \toolname{}, where our tool is about $\mathbf{1.45x}$ faster than Flie.

Even though Flie and our tool both encode as a SAT problem, Flie guarantees minimality by gradually increasing the size of the learned formula and invoking a SAT solver for each bound. However, we guarantee minimality by leveraging MaxSAT solving, which also involves solving multiple SAT instances but with the capability of sharing intermediate results (e.g., conflicting clauses) \cite{handbook-maxsat}.



\noindent\textbf{RQ2.} Figure \ref{fig:comparison}b shows the results for constrained problems, benchmarked against Flie. We find that our tool is faster than or equal to Flie (including where both time out) in most of the problems, specifically $\mathbf{30/30}$ in Peterson's algorithm, $\mathbf{10/10}$ in voting machine, $\mathbf{20/20}$ in specification repair, and $\mathbf{71/78 \approx 91\%}$ in invariant weakening.

Moreover, Table \ref{tab:compare-ltlsketch} shows the number of solved problems and the average solving time for Flie, LTLSketcher, and \toolname{}. Our tool solves the most number of problems and also has the lowest average solving time. On average, we solve about \textbf{3.2x} as many problems as Flie and \textbf{8.7x} as many as LTLSketcher. Also, our tool is about \textbf{3x} faster than both Flie and LTLSketcher.

\edit{Although \toolname{} also failed to solve some problems due to timeout,} \remove{The primary reason for our performance improvement is that}\edit{it is better than Flie and LTLSketcher in that:} Flie needs to enumerate \edit{a large, unconstrained search space} \remove{a large number of solutions }which \remove{often }leads to timeout \edit{more often}. \edit{LTLSketcher failed due to the lack of expressiveness (in defining the expected constraints) and enumeration capability.} \remove{LTLSketcher does not outperform both Flie and our tool because: (1) it fails to find the expected formulas due to the lack of expressiveness in defining constraints, and (2) it does not support enumeration. }Thus, we show that strong expressive power and the enumeration capability are both critical to LTL learning, and our tool supports both. \edit{We can solve all problems that Flie and LTLSketcher can without the loss of generality and much performance overhead.}

\subsection{Threats to Validity}
As shown in RQ2, Flie and LTLSketcher are outperformed in the constrained LTL problem benchmark. This might be due to the fact that our benchmark is based on the  set of use cases that we manually constructed to demonstrate the applicability of constrained LTL learning.
However, we believe that this potential bias is mitigated, as these problems contain a wide range of FOL constraints and are not specific to our Alloy-based encoding. We also note that the solutions found by \toolname{} are also in the space of possible solutions that Flie and LTLSketcher can generate.




\section{Related Work}
Neider and Gavran \cite{Neider2018-Learning} present Flie, a SAT-based encoding for LTL learning, which is the first approach of learning any unrestricted minimal LTL formulas over infinite traces.
It also presents a decision-tree based learning algorithm which is more efficient but does not guarantee the formula to be minimal. SCARLET \cite{Raha_scalable_2022} was proposed to overcome the scalability limitations of \cite{Neider2018-Learning}; however, they deal with a strict fragment of LTL
and also do not provide guarantees on minimality.

\begin{table}[!t]
\setlength\tabcolsep{1.8pt}
\caption{Comparisons of the number of solved problems and the average solving time (including the 180s timeout).}
\label{tab:compare-ltlsketch}
\resizebox{\linewidth}{!}{
\begin{tabular}{crrrrrrrr}
\toprule
\multicolumn{1}{l}{} & \multicolumn{2}{c}{Peterson's} & \multicolumn{2}{c}{Voting} & \multicolumn{2}{c}{Repair} & \multicolumn{2}{c}{Weakening} \\ \cmidrule(lr){2-3} \cmidrule(lr){4-5} \cmidrule(lr){6-7} \cmidrule(lr){8-9}
\multicolumn{1}{l}{} & \multicolumn{1}{c}{\#Solved} & \multicolumn{1}{c}{Time} & \multicolumn{1}{c}{\#Solved} & \multicolumn{1}{c}{Time} & \multicolumn{1}{c}{\#Solved} & \multicolumn{1}{c}{Time} & \multicolumn{1}{c}{\#Solved} & \multicolumn{1}{c}{Time} \\ \midrule
ATLAS & \textbf{30/30} & \textbf{43.65s} & \textbf{10/10} & \textbf{3.08s} & \textbf{20/20} & \textbf{37.25s} & \textbf{53/78} & \textbf{66.75s} \\
Flie & 2/30 & 179.68s & 2/10 & 144.92s & 1/20 & 178.63s & 30/78 & 142.98s \\
LTLSketcher & 0/30 & 180s & 2/10 & 144.30s & 0/20 & 180s & 11/78 & 160.73s \\
\bottomrule
\end{tabular}
}
\vspace{-10pt}
\end{table}

Before Neider's approach, techniques for mining LTL, such as \cite{Li2011-ltl-mining,Lemieux2015-ltl-mining,Ernst07-mining}, rely on patterns or templates. For example, given a template $\mathbf{G}(x \rightarrow \mathbf{F} y)$, Texada \cite{Lemieux2015-ltl-mining} infers specifications by substituting $x$ and $y$ with atomic events from the given traces (and does not learn from negative traces). Compared to our approach, Texada lies on a different trade-off point between expressiveness and efficiency: Although \toolname{} gives the user more control over the properties of LTL expressions to be learned, Texada is able to mine from much larger traces (e.g., thousands of events). 
More recently, Lutz et al. present LTLSketcher \cite{Lutz2023-LTL-Sketch}, where they consider three types of substitutions (i.e., any formula, unary operator, and binary operator) and propose a SAT-based method to find valid substitutions. Although they have improved expressiveness compared to Texada, they cannot express many patterns (e.g., CNF and DNF) as demonstrated in our use cases. In addition, none of these tools supports custom optimization objectives.

Researchers have also explored different variations of the learning problem~\cite{neider_expanding_2022}.
Gaglione et al. \cite{gaglione_maxsat-based_2022} present an approach using MaxSAT to learn from noisy data.
Although \alloymax{} also uses MaxSAT, we cannot solve this type of problem because \alloymax{} cannot directly control the weights of clauses in a MaxSAT instance.
Roy et al. \cite{roy2023learning} present an approach for learning from positive-only examples. While our tool also supports such a problem, we can only constrain and optimize  the syntactic structure of a solution whereas they can guarantee semantic minimality (i.e., a minimal set of behavior).




\section{Conclusion and Future Work}

We have proposed the \emph{constrained LTL learning problem} as a generalization of LTL learning from examples, and demonstrated the applicability and performance of \toolname{}, our prototype implementation tool. While our evaluation shows promise, we plan to investigate further ways to improve the learning method. First, the scalability of our tool is limited by the underlying MaxSAT solver, and an interesting future work is to investigate an approach that uses a different learning method (based on machine learning, for example). We also plan to investigate methods or heuristics for further reducing the space of possible LTL solutions (e.g., an encoding scheme that eliminates redundant, equivalent LTL expressions). \remove{Finally}\edit{Moreover}, we plan to investigate additional use cases for \toolname{} (such as invariant synthesis for distributed systems).

Our approach improves the expressive power of LTL learning tools. However, crafting constraints in FOL can be challenging for non-experts in formal logic. As future work, we plan to improve the usability of LTL learning by building on \toolname{} (e.g., through a DSL front-end or an LLM that converts natural language to FOL). In addition, \toolname{} and other learning tools require the user to provide examples. Future work could explore integrating external methods (e.g., model checkers or test generation tools) to (semi-)automate the generation of examples.

\section*{Acknowledgment}

This work was supported in part by the NSF awards 2144860 and 2319317, and the NSA grant H98230-23-C-0274. Any views, opinions, findings and conclusions or recommendations expressed in this material are those of the author(s) and do not necessarily reflect the views of the organizations.

\section*{Data Availability}
The source code of our tool and all the experimental results are available at: https://doi.org/10.5281/zenodo.14578202

\bibliographystyle{IEEEtran}
\bibliography{IEEEabrv,ref}

\clearpage
\section*{Appendix}

\subsection{\camera{Correctness Proof}}
\camera{
Our approach is sound but complete only up to the bound of the user-provided max number of DAG nodes.
\begin{theorem}[Soundness]
Given a constrained learning problem $\langle AP, \mathcal{S}, \Phi, \Psi \rangle$ and an upper bound of nodes $n$, a solution to the translated \alloymax{} problem is an LTL formula that is consistent with $\mathcal{S}$, satisfies $\Phi$, and optimizes against $\Psi$.
\end{theorem}
\begin{theorem}[Completeness]
If there is a valid solution to a constrained learning problem within a bound $n$, our approach is guaranteed to find it.
\end{theorem}
}

\camera{
\textit{Proof.} For soundness, recall that  input traces are in the lasso form (i.e., $uv^{\omega}$). As demonstrated in Figure \ref{fig:lasso-trace}, although a lasso is infinite, the set of all states that are reachable from any time $t$ is finite and can be computed from the finite prefix $uv$. Thus, the valuation of an LTL formula on a lasso can be decided based only on $uv$ \cite{Neider2018-Learning}. Our semantic encoding is inspired by the established encoding of SAT-based LTL model checking \cite{Neider2018-Learning,Handbook-model-checking-SAT}, and the correctness of the \alloymax{} to MaxSAT translation is established in \cite{Alloy-Max}. For completeness, the syntax encoding in Section~\ref{sec:syntax-encoding} guarantees that it will search for all possible formulas given the bound. Moreover, the objective $\Psi$ does not change the amount of valid solutions w.r.t. sample $\mathcal{S}$ and constraint $\Phi$ \cite{Alloy-Max}. For instance, with objective $\mathcal{L} \cup \mathcal{R} \approx \emptyset$, the solver first returns a formula of minimal size and can enumerate all valid solutions within the bound in ascending order of formula size. $\hfill\square$
}

\camera{
Specifically, the correctness of the above proof relies on the following theorems.
\begin{theorem}
For any LTL formula $\phi$, its valuation on a lasso-shaped trace $uv^\omega$ can be decided purely on the finite prefix $uv$.
\end{theorem}
}

\camera{
\textit{Proof.} In this work, we consider only future-time operators (i.e., $\textbf{X, U, F, G}$). For example, consider the next operator $\textbf{X}$, given an LTL formula $\textbf{X}\phi$ and a lasso trace $uv^\omega$, the valuation of $uv^\omega, i \models \textbf{X}\phi$ depends on the valuation of $uv^\omega, i+1 \models \phi$. In the case where $i+1 > |uv|$, the trace cycles within $v$, and thus the valuation can be reduced to $uv^\omega, |u| + \bigl((i + 1 - |u|) \mod |v|\bigr) \models \phi$. The same reduction can be derived for other temporal operators \cite{Neider2018-Learning}. $\hfill\square$
}

\camera{
Therefore, for any valuation $uv^\omega, i \models \phi$, we only need to compute the finite set of future states from the time $i$. We then show that our Alloy encoding can compute the future states of any lasso trace and any time point.
\begin{theorem}
For a lasso trace $uv^\omega$ and a time point $i$, our Alloy encoding correctly computes the finite set of future states of $uv^\omega[i, \infty]$.
\end{theorem}
}

\camera{
\textit{Proof.} The computation of future states is based on the relations $next$ and $lasso$. For example, for the trace $t$ described in Figure \ref{fig:lasso-trace}, we have $next = \{(0, 1), (1, 2)\}$ and $t.lasso = \{(2, 1)\}$. Thus, $next + t.lasso = \{(0, 1), (1, 2), (2, 1)\}$, and the expression $i.(next + t.lasso)$ computes the next time point (state) of time $i$ with the consideration of the cycle (e.g., when $i = 2$, the expression evaluates to $\{1\}$). Moreover, for the $\textbf{F}, \textbf{G},~\text{and}~\textbf{U}$ operators, we use the $\textsf{futureIdx}(t, i)$ function to compute the set of all future time points (states) of a trace $t$ starting from time $i$. It is defined as follows:
}

\begin{alloy}
fun futureIdx[t: Trace, i: SeqIdx]: set SeqIdx {
  i.^(next + t.lasso) + i
}
\end{alloy}

\camera{The $^\wedge$ operator computes the transitive closure of a relation. Thus, in our example, $^\wedge(next + t.lasso) = \{(0, 1), (0, 2),$ $(1, 2), (1, 1), (2, 1), (2, 2)\}$, and the expression $i.^\wedge(next + t.lasso)$ returns the set of future states from time $i$ (e.g., when $i = 0$, it evaluates to $\{1, 2\}$; and when $i = 2$, it evaluates to $\{1, 2\}$ where both $1$ and $2$ are returned because of the cycle). In addition, we append the current time point $i$ to the set based on the LTL semantics. Finally, this set of future states are used to evaluate an LTL expression. $\hfill\square$}

\camera{
Finally, we show our Alloy model encodes the LTL semantics.
\begin{theorem}
The Alloy encoding has the same semantics as the semantics of LTL.    
\end{theorem}
}

\camera{
\textit{Proof.} It is straightforward to see that our Alloy encoding models the semantics of LTL as both of them are based on first-order logic expressions. For example, the interpretation of $uv^\omega, i \models \mathbf{F}\phi$ is defined as $\exists j: i \leq j \land uv^\omega, j \models \phi$. This is represented in our Alloy encoding as:}
\begin{alloy}
some j: futureIdx[t, i] | n.l->j in t.val
\end{alloy}
\camera{
where \textsf{some} corresponds to $\exists$, $t$ is the trace, \textsf{futureIdx} computes the set of future states from $i$, $n$ is the expression $\mathbf{F}\phi$, and $n.l$ is the sub-expression $\phi$. One exception is the encoding of the $\textbf{U}$-operator, where we leverage the heuristics from bounded model checking to unfold it using the $\textbf{X}$-operator, i.e., $q \mathbf{U} p = p \lor (q \land \mathbf{X}(q \mathbf{U} p))$. This heuristic can produce a more succinct encoding and thus improve the performance. $\hfill\square$
}

\end{document}